\documentclass[pra,twocolumn,aps,amsmath]{revtex4}

\usepackage{graphics}
\usepackage{epsfig}
\usepackage{float}
\usepackage{subeqn}
\usepackage{subfigure}
\usepackage[english]{babel}

\begin{document}
\newcommand*{\ket}[1]{$|{#1}\rangle$}
\newcommand*{\beq}{\begin{equation}}
\newcommand*{\eeq}{\end{equation}}
\newcommand*{\beqs}{\begin{equation*}}
\newcommand*{\eeqs}{\end{equation*}}
\newcommand{\duh}[2]{\left(1+\frac{z_{#1#2}}{r_{#1}}\right)^2+\left(\frac{\lambda z_{#1#2}}{w_{#1} \ell_{#1}}\right)^2}
\newcommand{\duhb}[2]{\left(1+\frac{z_{#1#2}}{r(z_{#1})}\right)^2+\left[\frac{\lambda z_{#1#2}}{s(z_{#1})
\ell(z_{#1})}\right]^2}
\newcommand{\gam}[2]{\sqrt{\duh{#1}{#2}}}

\title{Gaussian Schell Source as Model for Slit-Collimated Atomic and Molecular Beams}
\author{Ben McMorran}
\author{Alexander D. Cronin}
 \affiliation{Department of Physics, University of Arizona, Tucson, AZ 85721}

\begin{abstract}
We derive the spatial coherence and intensity profiles of beams emerging from two consecutive collimating apertures, and compare our results with data. We show how to make a Gaussian Schell-model (GSM) beam by assuming Gaussian apertures. Then we compare the intensity profile, the transverse coherence width and the divergence angle of a GSM beam with those same properties of a beam that is collimated with two hard-edged slits. The GSM beam formulae are simpler, and offer an intuitive way to understand how partially coherent beams interact with interferometers.
\end{abstract}

\date{\today}
\maketitle

\section{Introduction}

\begin{figure}
\includegraphics[width = 8cm]{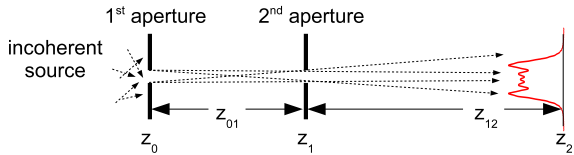}
\caption{In this paper, we assume that an aperture with characteristic width $s_0$ (either a slit or a Gaussian filter) placed at $z=z_0$ is incoherently illuminated. A second similar aperture with width $s_1$ is placed a distance $z_{01}$ away. Generally, beams resulting from this arrangement are \emph{partially} coherent; the characteristic size of the mutual coherence function (the coherence width) is less than or equal to the width of the intensity distribution.}\label{fig:setup}
\end{figure}

Perhaps the simplest way to form a beam of particles is to place two consecutive apertures in front of a source. As illustrated in Figure \ref{fig:setup}, a first aperture placed directly in front of an extended incoherent source helps define the source. Then a second aperture, farther away, collimates the beam. In many experiments with X-rays, gamma rays, atoms, molecules, neutrons or electrons a simple ray picture can adequately predict the intensity profile of such beams. But more care is needed to model the spatial coherence that is so crucial for understanding the behavior of interferometers.

A Gaussian Schell-model (GSM) beam is a particular type of partially coherent wave field that is often used to model beam coherence properties (transverse coherence lengths and wavefront radius of curvature) as well as intensity profiles.  In fact, we have learned quite a bit about three-grating interferometers with GSM beams \cite{MCC07}.  The major benefit of GSM beams is that the Huygens-Fresnel propagation can be described analytically in many cases.  However, the question has been raised, `Is the GSM a realistic way to model the beams in matter wave interferometers?'  The purpose of this paper is to answer this question in the affirmative.  We show how to make a GSM beam and then we describe how the GSM beam's properties compare to a slit-collimated beam made with hard-edged slits.


Furthermore, we show how a simple Gaussian Schell-model (GSM) can be used to accurately predict the characteristic width, the transverse coherence length (coherence width), and the divergence angle of a slit-collimated beam. In \cite{MCC07}, we show how a GSM beam can be used to efficiently analyze the role of partial coherence in a variety of grating interferometers, and we suggest this model can now be applied to several matter wave interferometry experiments. To this end, the purpose of this manuscript is to show that GSM beams can accurately model slit-collimated beams commonly used for matter wave experiments.

\begin{figure}
\includegraphics[height=5cm]{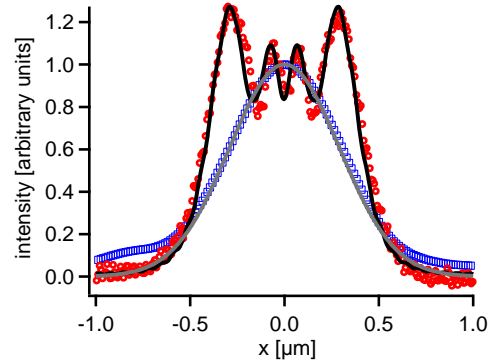}
\caption{Actual intensity profile of a beam emanating from collimating hard-edged slits (circles) and Gaussian apertures (squares), compared to theory (solid lines) developed using partially coherent optics - Equations \ref{eq:J2_slit} and \ref{eq:J2_gsm}. For both sets of data, an orange LED ($\lambda \approx$ 600 nm) was placed directly in front of a millimeter-wide slit, which illuminated a second slit of the same width 1 meter away. Intensity profiles were obtained using a CCD placed 10 cm behind the second slit.}
\label{fig:profilefit}
\end{figure}

\section{Theory - Partially coherent beams from two consecutive apertures}

Here we introduce the general formalism we will use. We first introduce the conventional form of a function that describes two-point correlations in a field. A semi-coherent scalar field can be described as a statistical distribution of coherent beams. One way to model this is to use the mutual intensity \cite{BOW59}, defined as
\begin{equation}
\label{eq:Jdefn} J(\pmb{\rho}_a,\pmb{\rho}_b) \equiv
\langle\psi(\pmb{\rho}_a)\psi^*(\pmb{\rho}_b)\rangle
\end{equation}
\noindent where the angle brackets denote an average over the ensemble. This expression is a way to quantify the correlation between two points, $\pmb{\rho}_a$ and $\pmb{\rho}_b$, in a particular plane transverse to the direction of propagation. This can be seen in an alternative form for the mutual intensity:
\begin{equation}
\label{eq:J_alt} J(\pmb{\rho}_a,\pmb{\rho}_b) = \sqrt{I(\pmb{\rho}_a)} \sqrt{I(\pmb{\rho}_p)}
\mu(\pmb{\rho}_a, \pmb{\rho}_b)
\end{equation}
\noindent where $\mu(\pmb{\rho}_a, \pmb{\rho}_b)$ is the spectral degree of coherence, also defined in \cite{BOW59}. The intensity of the light field at a point $\pmb{\rho}$ can be found by setting the two points of the mutual intensity equal to eachother:
\begin{equation}
\label{eq:IfromJ} I(\pmb{\rho})=J(\pmb{\rho},\pmb{\rho})
\end{equation}
\noindent For use with polychromatic fields, such as those commonly encountered in matter wave beams, the mutual intensity may also be used to calculate the spectral density, as noted by Wolf and Devaney \cite{WOD81}.

The mutual intensity provides a way to use techniques developed for coherent optics to examine the evolution of semi-coherent optical fields. The difference is that the semi-coherent calculation requires two operations (one for each of the two correlated points) whereas the conventional coherent calculation requires only one. For example, when calculating how a thin optical element modifies a mutual intensity distribution, the complex transmission function $t(\pmb{\rho})$ describing the element appears twice:
\begin{eqnarray}
\label{eq:Joptic} J(\pmb{\rho}_{a},\pmb{\rho}_{b};z^+) &=&t(\pmb{\rho}_{a})t^*(\pmb{\rho}_{b})J(\pmb{\rho}_{a},\pmb{\rho}_{b};z)
\end{eqnarray}
\noindent where $z^+$ denotes a position just after the element. Likewise, one can use the paraxial (Fresnel) approximation to the Huygens-Fresnel principle to propagate a partially coherent wave from a source plane $z_0$ over a distance $z_{01}$ to plane $z_1$ \cite{FRS82}:
\begin{eqnarray}
\label{eq:Jprop} J(\pmb{\rho}_a,\pmb{\rho}_b;z_1) &=& \frac{1}{\lambda^2 z_{01}^2}\int\!\!\!\int  e^{-\frac{i\pi}{\lambda
z_{01}}\left[(\pmb{\rho}_a-\pmb{\rho}'_a)^2-(\pmb{\rho}_b-\pmb{\rho}'_b)^2\right]} \nonumber\\
&& \times J(\pmb{\rho}'_a,\pmb{\rho}'_b;z_0) d\pmb{\rho}'_a d\pmb{\rho}'_b.
\end{eqnarray}
\noindent where $\pmb{\rho}'_a$ and $\pmb{\rho}'_b$ are points in the source plane. The Fresnel propagator we have chosen to use - the quadratic phase factor - approximates hemispherical Huygens wavelets with paraboloids, and it is valid for calculating fields not just far from the source plane but also near to it. Equation \ref{eq:Jprop} is \emph{not} the van Cittert-Zernike theorem, which is restricted for use with fairly incoherent sources. It is rather a paraxial approximation to Zernike's more general propagation law for fields with arbitrary coherence properties \cite{ZER38}.

Just as in coherent optics, the propagation of partially coherent fields through complex optical systems may be computed using succesive applications of Equations \ref{eq:Joptic} and \ref{eq:Jprop}. For the remainder of this manuscript, we will only evaluate the beam's mutual intensity in a single transverse direction.

\subsection{`Hard-edge slits'}

Let us now use the above framework to calculate the partially coherent beam emerging from the two aperture system illustrated in Figure 1, assuming slits with hard edges. We will only calculate the mutual intenstiy in one transverse direction $\hat{x}$. The complex transmission function for the first slit with hard edges is

\begin{equation}
\label{eq:t0_slit} t_0(x) = \textrm{rect}\left(\frac{x}{s_0}\right)
\end{equation}

\noindent where $s_0$ is the width of the slit and, as in \cite{GAS78}, the rect funtion is defined as

\begin{equation}
\label{eq:rect} \textrm{rect}\left(\frac{x}{s}\right) \equiv \left\{ \begin{array}{ll}
0 \textrm{,} & \textrm{$|x|>s/2$}\\
\frac{1}{2} \textrm{,} & \textrm{$|x|=s/2$}\\
1 \textrm{,} & \textrm{$|x|<s/2$}
\end{array} \right.
\end{equation}

\noindent When this aperture is placed in front of a completely incoherent source with radiant emittance $I_0$, at $z_0$, the mutual intensity directly behind the slit, at $z_0^+$, is given by

\begin{equation}
\label{eq:J0_slit} J(x_a,x_b;z_0^+) = I_0 \textrm{rect}\left(\frac{x_a}{s_0}\right)\textrm{rect}\left(\frac{x_b}{s_0}\right)\delta(x_b - x_a).
\end{equation}

\noindent The incoherent nature of the source is reflected in the presence of the delta function; i.e. there is no correlation between any two points.

The 1-D form of Equation \ref{eq:Jprop} is used to calculate the field a distance $z_{01}$ away. Making use of the delta function in Equation \ref{eq:J0_slit}, and recognizing that $\textrm{rect}(x)^2=\textrm{rect}(x)$ we find:

\begin{eqnarray}
\label{eq:J1_slit_a} J(x_a,x_b;z_1) &=& \frac{I_0}{\lambda z_{01}}\int\!\!\!\int  e^{-\frac{i\pi}{\lambda
z_{01}}\left[(x_a-x'_a)^2-(x_b-x'_a)^2\right]} \nonumber\\
&& \times \textrm{rect}\left(\frac{x_a}{s_0}\right) dx'_a
\end{eqnarray}

\noindent The remaining integral over $x'_a$ becomes a straightforward Fourier transform of a rect function and we arrive at:

\begin{eqnarray}
\label{eq:J1_slit} J(x_a,x_b;z_1) &=& I_1 \textrm{sinc}\left[\frac{s_0 (x_b-x_a)}{\lambda z_{01}}\right] e^{\frac{i\pi}{\lambda
z_{01}}\left(x_b^2-x_a^2\right)}
\end{eqnarray}

\noindent where $I_1 = I_0 s_0/\lambda z_{01}$ and

\begin{equation}
\label{eq:sinc} \textrm{sinc}(ax) \equiv \frac{\sin(\pi a x)}{\pi a x}.
\end{equation}

We now let this mutual intensity pass through a second aperture, another hard-edged slit of width $s_1$:

\begin{eqnarray}
\label{eq:t1_slit} t_1(x) &=& \textrm{rect}\left(\frac{x}{s_1}\right)
\end{eqnarray}

\noindent The mutual intensity just after this aperture (denoted by $z_1^+$) becomes:

\begin{eqnarray}
\label{eq:J1+_slit} J(x_a,x_b;z_1^+) &=& I_1 \textrm{rect}\left(\frac{x_a}{s_1}\right)\textrm{rect}\left(\frac{x_b}{s_1}\right) \\
&& \times \textrm{sinc}\left[\frac{s_0 (x_b-x_a)}{\lambda z_{01}}\right] e^{\frac{i\pi}{\lambda
z_{01}}\left(x_b^2-x_a^2\right)} \nonumber
\end{eqnarray}

Now we wish to determine the properties of the beam an arbitrary distance away from the second collimating slit. We may again apply the paraxial approximation to Zernike's propagation law (Equation \ref{eq:Jprop}):

\begin{eqnarray}
\label{eq:J2_slit} J(x_a,x_b;z_2)
%
&=& \frac{I_1}{\lambda z_{12}}e^{\frac{i\pi}{\lambda
z_{12}}(x_b^2-x_a^2)}\int\!\!\!\int e^{-\frac{i2\pi}{\lambda
z_{12}}(x'_b x_b - x'_a x_a)} \nonumber\\
&& \textrm{rect}\left(\frac{x'_a}{s_1}\right)\textrm{rect}\left(\frac{x'_b}{s_1}\right) \textrm{sinc}\left[\frac{s_0 (x'_b-x'_a)}{\lambda z_{01}}\right]   \nonumber\\
&& e^{\frac{i\pi}{\lambda}\left({x'_b}^2-{x'_a}^2\right)\left(\frac{1}{z_{01}}+\frac{1}{z_{12}}\right)} dx'_a dx'_b.
\end{eqnarray}

\noindent Even after rearranging terms to reveal a two dimensional Fourier transform, it is difficult to spot a full analytical solution to Equation \ref{eq:J2_slit}. This is unfortunate since in many optical and matter wave experiments great insight could be gained by analytically computing how such a beam evolves through a given system. It is for this reason that we suggest using a GSM beam to model these experiments. The Fourier transforms present in Equation \ref{eq:J2_slit} are conducive to numerical evaluation by a computer, however, and later in the manuscript we will use this technique to compare this slit-collimated beam to a GSM beam.

\subsection{Gaussian apertures - GSM beams}

Again, we now calculate the partially coherent beam emerging from the system illustrated in Figure \ref{fig:setup}, but this time assuming the two Gaussian apertures. In order to compare the dimensions of the resulting GSM beam to those of the slit-collimated beam, we must first choose the size of the Gaussian apertures carefully. The transmission function of the first aperture is

\begin{eqnarray}
\label{eq:t0_gauss} t_0(x) &=& e^{\frac{-\pi x^2}{s_0^2}}
\end{eqnarray}

\noindent where we choose a convention for the width $s_0$, which is just slightly larger than the FWHM, such that the total area of the Gaussian aperture is equal the the total area of a hard-edged slit with the same width \cite{GAS78}.


The mutual intensity directly behind this incoherently illuminated Gaussian aperture is given by

\begin{equation}
\label{eq:J0_gsm} J(x_a,x_b;z_0^+) = I_0 e^{\frac{-\pi}{s_0^2}(x_a^2 + x_b^2)}
\delta(x_b - x_a).
\end{equation}

\noindent As before, the mutual coherence function after this first aperture is a delta function, indicating there is no correlation between any two points.

Like in the previous section, we now insert Equation \ref{eq:J0_gsm} into the 1-D form of Equation \ref{eq:Jprop} to calculate the field emitted by this Gaussian source a distance $z_{01}$ away:

\begin{eqnarray}
\label{eq:J1_gsm_a} J(x_a,x_b;z_1) &=& \frac{I_0}{\lambda z_{01}}\int\!\!\!\int  e^{-\frac{i\pi}{\lambda
z_{01}}\left[(x_a-x'_a)^2-(x_b-x'_b)^2\right]} \nonumber\\
&& \times e^{\frac{-\pi}{s_0^2}({x'_a}^2 + {x'_b}^2)}
\delta(x'_b - x'_a) dx'_a dx'_b
\end{eqnarray}

\noindent After making use of the delta function by integrating over $x'_b$, the remaining integral becomes a straightforward Fourier transform of a Gaussian function and we arrive at:

\begin{eqnarray}
\label{eq:J1_gsm} J(x_a,x_b;z_1) &=& I_1 e^{\frac{-\pi}{2\sigma_1^2}(x_b-x_a)^2} e^{\frac{i\pi}{\lambda
r_1}\left(x_b^2-x_a^2\right)}
\end{eqnarray}

\noindent where $I_1 = I_0 s_0/\lambda z_{01}$,

\begin{eqnarray}
\label{eq:r} r_1 &=& z_{01}
\end{eqnarray}

\noindent is the radius of wavefront curvature, and

\begin{eqnarray}
\label{eq:sigma} \sigma_1 &=& \lambda z_{01}/s_0
\end{eqnarray}

\noindent is the correlation width. We will discuss the physical meaning of the correlation width shortly.

The field described by Equation \ref{eq:J1_gsm} is incident upon a second Gaussian-weighted aperture with width $s_1$ located at $z_1$:

\begin{eqnarray}
\label{eq:t1_gsm} t_1(x) &=& e^{\frac{-\pi x^2}{s_1^2}}
\end{eqnarray}

\noindent The mutual intensity just after this second aperture (denoted by $z_1^+$) can be found using Equation \ref{eq:Joptic}:

\begin{eqnarray}
\label{eq:J1+_gsm} J(x_a,x_b;z_1^+) &=& t_1(x_a)t^*_1(x_b)J(x_a,x_b;z_1) \\
&=& I_1 e^{\frac{-\pi }{s_1^2}(x_a^2+x_b^2)} e^{\frac{-\pi}{2\sigma_1^2}(x_b-x_a)^2} e^{\frac{i\pi}{\lambda
r_1}\left(x_b^2-x_a^2\right)} \nonumber.
\end{eqnarray}

\noindent We now define the \emph{transverse coherence length} or \emph{coherence width} $\ell_1$

\begin{eqnarray}
\label{eq:ell} \frac{1}{\ell_1^2} \equiv \frac{1}{\sigma_1^2} + \frac{1}{s_1^2}
\end{eqnarray}

\noindent so that we may write the partially coherent beam exiting from the second aperture in a more recognizable form:

\begin{eqnarray}
\label{eq:J1+_gsm_a} J(x_a,x_b;z_1^+) &=& I_1 e^{\frac{-\pi }{2 s_1^2}(x_a+x_b)^2} e^{\frac{-\pi}{2 \ell_1^2}(x_b-x_a)^2} e^{\frac{i\pi}{\lambda
r_1}\left(x_b^2-x_a^2\right)}.
\end{eqnarray}

\noindent Equation \ref{eq:J1+_gsm_a} is the usual representation of a Gaussian Schell-model (GSM) beam \cite{FRS82,MAW95}. The GSM beam can be thought of as an ensemble of coherent beams each with an average momentum and radius of curvature that depends on its displacement from the optical axis.

The correlation width $\sigma_1$ is commonly touted as the characteristic maximum distance between any two points in plane $z_1$ for which there is significant correlation in the field \cite{GOR05}. However, this statement loses its meaning when the correlation width exceeds the size of the classical intensity distribution (a feat easily accomplished by choosing a sufficiently small width for the first slit or second slit). A hypothetical Young's experiment, which is the most direct method for measuring two point correlations, with slit separations much greater than the extent of the beam would not yield visible interference fringes.

The coherence width $\ell_1$ is more directly related to experiment, as it can never be larger than the transverse extent of the beam. The coherence width is the maximum separation between slits in a Young's experiment for which one can expect to see large intensity modulations in the fringe pattern. Thus, we elect to use Equation \ref{eq:J1+_gsm} as the form for the GSM beam.

Upon another application of Equation \ref{eq:Jprop}, one may derive an analytical solution for the GSM beam at any distance from the second aperture:
\begin{eqnarray}
\label{eq:J2_gsm} J(x_a,x_b;z_2) &=& I(z_2) e^{-\pi\left[\frac{(x_a+x_b)^2}{2s(z_2)^2} + \frac{(x_b-x_a)^2}{2\ell(z_2)^2} + \frac{i\left(x_b^2-x_a^2\right)}{\lambda
r(z_2)}\right]}
\end{eqnarray}
\noindent where $I(z_2) \equiv I_0 s_0/\lambda z_{02}$ and
\begin{eqnarray}
s(z_2) = s_1 \gam{1}{2}, \label{eq:w_z} \\
\ell(z_2) = \ell_1 \gam{1}{2}, \label{eq:ell_z} \\
r(z_2) = z_{12}\left[
\frac{\duh{1}{2}}{\frac{z_{12}}{r_{1}}\left(1+\frac{z_{12}}{r_{1}}\right)+\left(\frac{\lambda
z_{12}}{s_1 \ell_1}\right)^2}\right]. \label{eq:r_z}
\end{eqnarray}

\section{Modelling a slit-collimated beam with a GSM beam}

\begin{figure}
\subfigure[  $\mathbf{z_{12} = 5 cm}$]{
\label{fig:atom_beam_profile:a} 
\includegraphics[height=5cm]{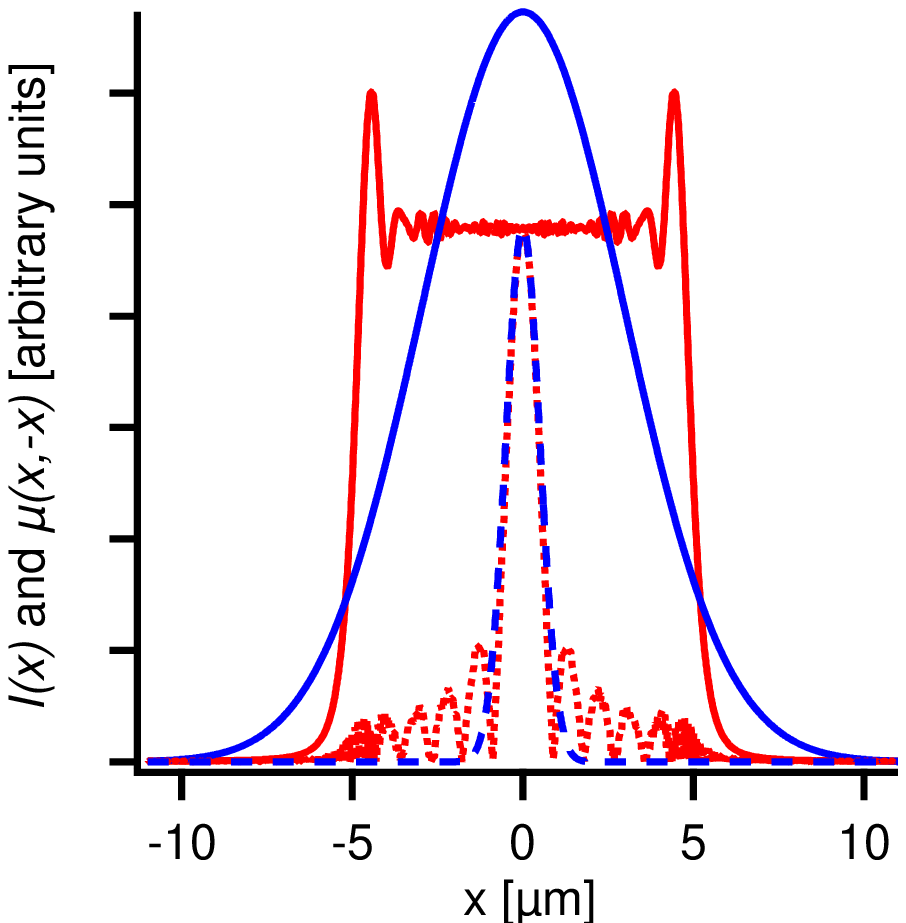}} \\
\vspace{0cm}
\subfigure[  $\mathbf{z_{12} = 2 m}$]{
\label{fig:atom_beam_profile:b} 
\includegraphics[height=5cm]{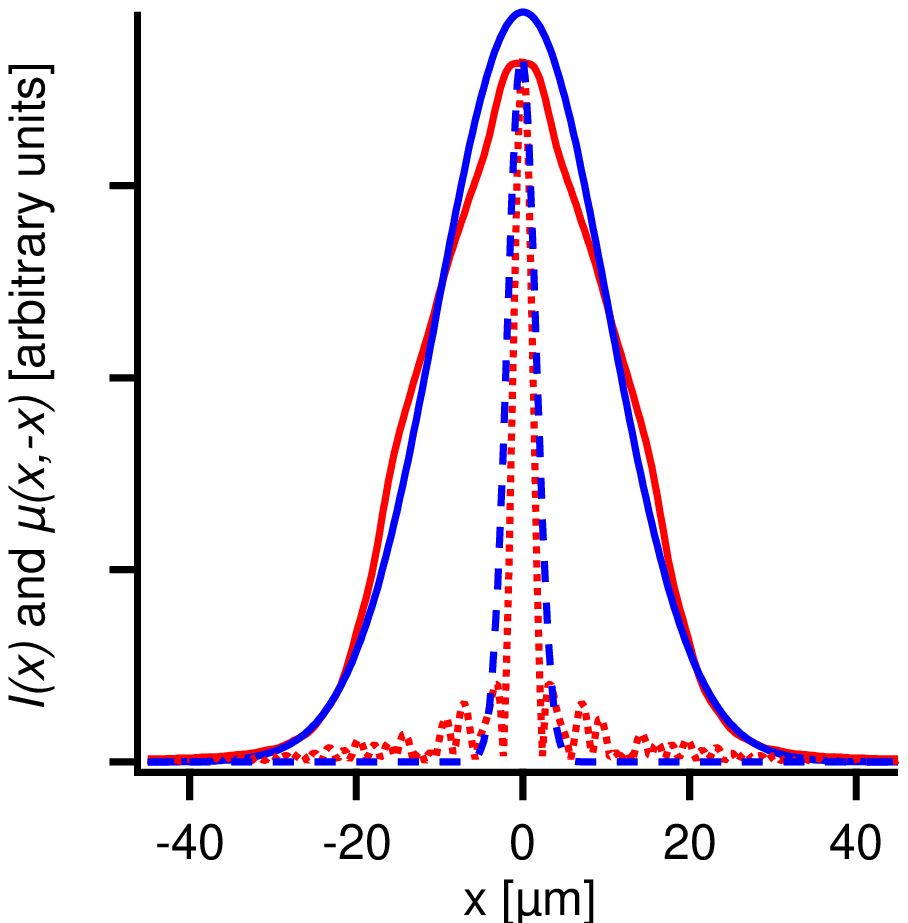}}
\caption{Comparison of slit-collimated and GSM beam intensities and mutual coherence profiles. Here we model our own sodium atom beam, using $\lambda =$ 17 pm for 1000 m/s atoms, and 10 $\mu$m slits spaced 1 meter apart. The intensity profile of the slit-collimated beam (solid red) was obtained numerically using Equation \ref{eq:J2_slit} and setting $x_a = x_b = x$. The mutual coherence profile at the center of the beam (dotted red) was obtained similarly, setting $x_a = -x_b$ and averaging over several wavelengths. The intensity profile of the GSM beam (solid blue) is just a Gaussian curve using Equation \ref{eq:w_z} for the width, and likewise for the mutual coherence profile (dotted blue) using Equation \ref{eq:ell_z}.}
\label{fig:atom_beam_profile}
\end{figure}

As stated previously, while Equation \ref{eq:J2_slit} provides a way to compute the mutual intensity of the partially coherent beam emerging from collimating slits exactly (within the limits of the Fresnel approximation), it is difficult to calculate how this beam interacts with any optical elements such as diffraction gratings. In matter wave interferometry, where one is often analyzing the occurence of interference fringes, knowledge about the detailed shape of the intensity distribution or mutual coherence function is often unnecessary. What is particularly useful, however, is knowledge of the characteristic size of these distributions - the intensity and spatial coherence widths, and radius of wavefront curvature. The GSM beam provides a much faster way to calculate these properties, and a simpler way to analytically propagate a partially coherent beam through a given optical system.

\begin{figure}
\subfigure[ slit-collimated $I(x;z_{12})$]{
\label{fig:all_beam:a} 
\includegraphics[height=3.5cm]{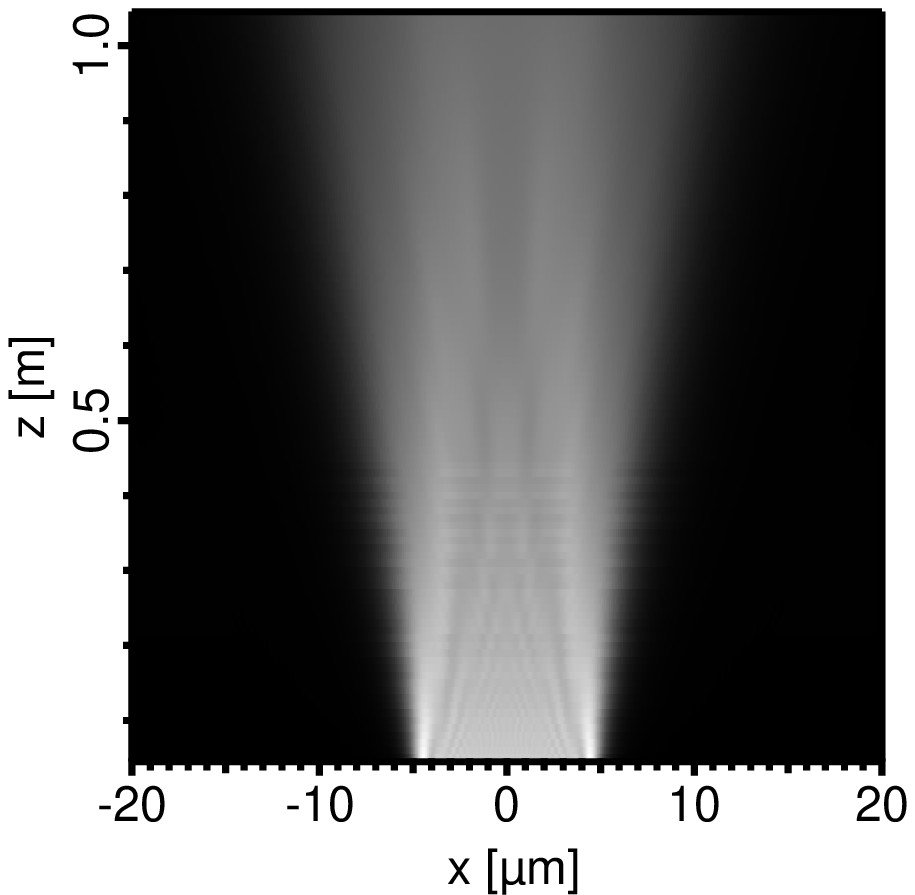}}
\subfigure[ slit-collimated $\mu(x,-x;z_{12})$]{
\label{fig:all_beam:b} 
\includegraphics[height=3.5cm]{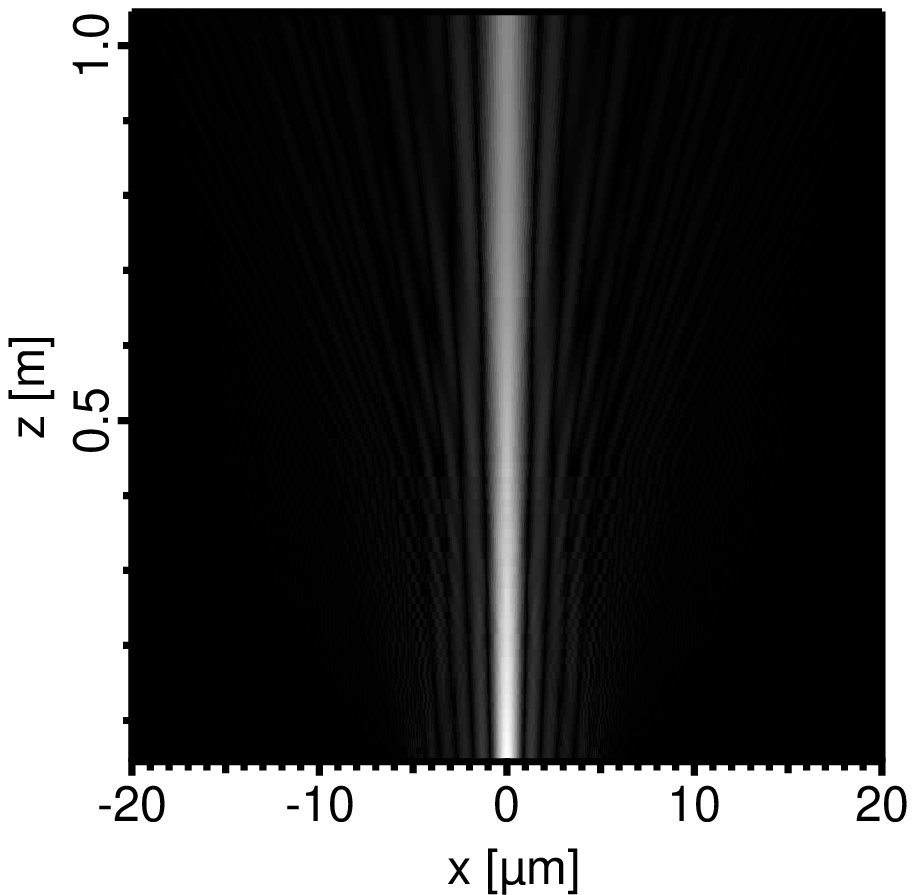}}\\
\vspace{0cm}
\subfigure[ GSM $I(x;z_{12})$]{
\label{fig:all_beam:c} 
\includegraphics[height=3.5cm]{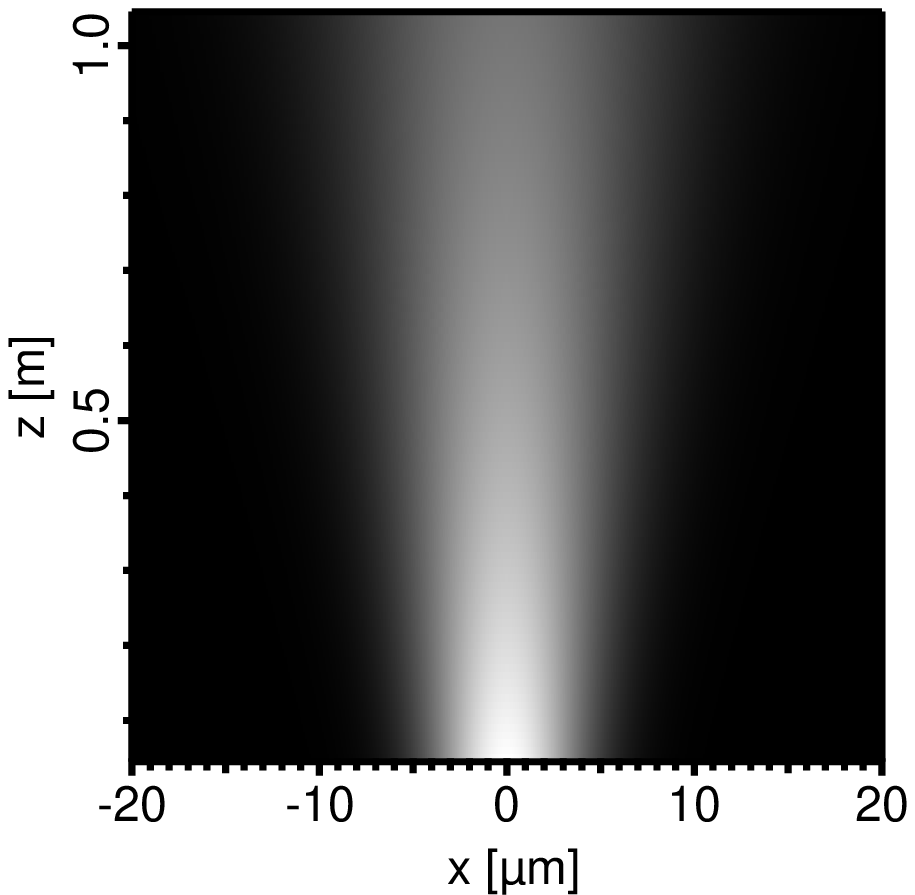}}
\subfigure[ GSM $\mu(x,-x;z_{12})$]{
\label{fig:all_beam:d} 
\includegraphics[height=3.5cm]{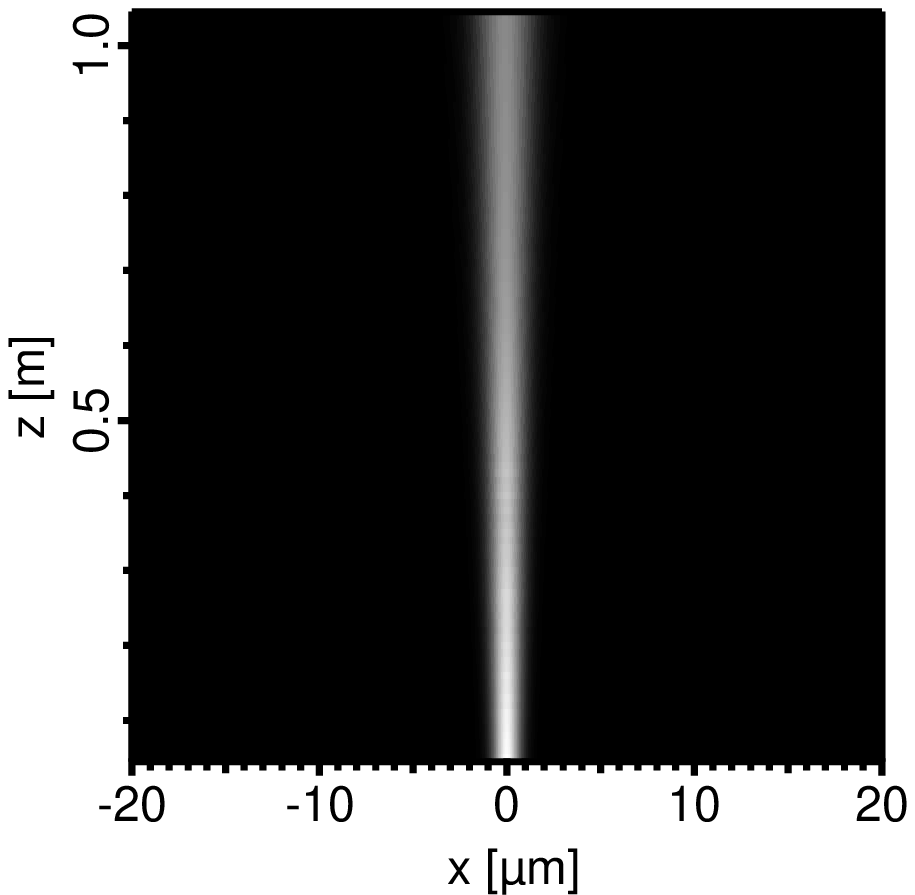}}
\caption{(a) Intensity profile and (b) mutual coherence profile of a beam emerging from a hard-edged collimating slit, as a function of the propagation distance. Comparable profiles (c) and (d) are shown for a GSM beam. These figures were generated with the same parameters used in Figure \ref{fig:atom_beam_profile}.}
\label{fig:all_beam}
\end{figure}

Comparing this field directly behind two consecutive slits to the field behind two consecutive Gaussian apertures (Figure \ref{fig:atom_beam_profile:a}), we of course see that the intensity distribution is entirely determined by the uniformly-illuminated second aperture. Note that the intensity of the GSM beam directly behind the second aperture is reduced to about 20\% of its maximum value at $x=\pm s_1/2$, whereas the intensity of the slit-collimated beam is reduced by half. When comparing profile widths for different types of beams we use the characteristic width of the profile when it is 20\% of its maximum.

%
%

\begin{figure}[H]
\centering
\includegraphics[width = 7cm]{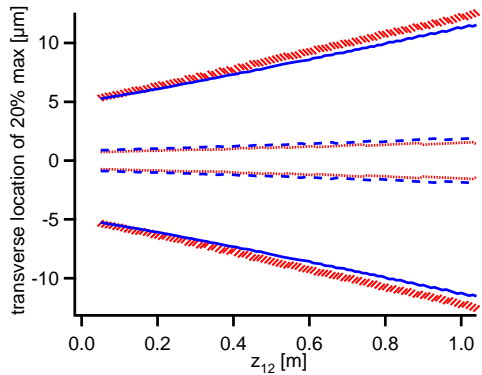}
\caption{Characteristic widths of the same beam profiles shown in Figure \ref{fig:all_beam}. The slashed red line displays the width of a hard-edged slit-collimated beam, the solid blue shows the width of a corresponding GSM beam, and the red dotted line and blue dashed line indicate the coherence width of the slit-collimated beam and the GSM beam, respectively (evaluated at the center of the beam).}\label{fig:beam_width_vs_dist}
\end{figure}

The ratio of transverse coherence length to beam width in Equations \ref{eq:w_z} and \ref{eq:ell_z} remains unchanged as the beam propagates through free space. That is,
\begin{equation}
\label{eq:beta} \beta \equiv \ell(z)/s(z) = \textrm{
constant}
\end{equation}
\noindent where $\beta$ is the `degree of global coherence' \cite{BOW59,MAW95}. If $\beta$ is equal to unity, Equations \ref{eq:w_z} to \ref{eq:r_z} describe the beam diameter and radius of wavefront curvature for a standard (fully-coherent) Gaussian beam. According to Figure \ref{fig:beam_width_vs_dist} the global degree of coherence of our sodium atom beam used for interferometry experiments is about $\beta = 0.1$.

\section{Conclusion}

In summary, we have shown that a Gaussian Schell-model (GSM) beam can efficiently model various properties of a slit-collimated beam. While the specific shapes of the intensity and coherence profiles can differ dramatically between the two types of beams, the characterstic size of these distributions are identical and evolve through space via propagation in a similar way. The primary advantage of using the GSM beam to model the complicated dynamics of a beam generated from hard-edged apertures is that it is much easier to analyze its propagation through various optical elements, such as those used in matter wave interferometry experiments. As an example of this, in an upcoming paper \cite{MCC07} we use a GSM beam to efficiently model a wide variety of grating interferometers commonly used in experiments, and the role that partial coherence and beam divergence has on their behavior.

This research was supported by the National Science Foundation Grant No. 0653623.

\bibliography{slitcollgsm}

\end{document}